\documentclass{PoS}
\usepackage{epsfig}
\PoS{PoS(LAT2005)328}
\newcommand{\beq}{\begin{equation}}
\newcommand{\eeq}{\end{equation}}

\title{Geometry of three dimensional vacuum domains in four dimensional SU(2)
gluodynamics}

\ShortTitle{Geometry of three dimensional vacuum domains in four
dimensional SU(2) gluodynamics. }

\author{\speaker{A.V. Kovalenko}, M.I.~Polikarpov, S.N.~Syritsyn
\\
        ITEP, B. Cheremushkinskaya 25, Moscow, 117259 Russia\\
        E-mail: \email{kovalenko@itep.ru}, \email{polykarp@itep.ru},
        \email{syritsyn@itep.ru}}
\author{V.I.~Zakharov\\
        MPI, F\"ohringer Ring 6, 80805, M\"unchen, Germany\\
        E-mail: \email{xxz.mppmu.mpg.de}}

\abstract{We review briefly recent results of lattice simulations on 3d
domains in the vacuum state of $SU(2)$ gluodynamics. The defects are
defined as unification of all the negative links in central projection
under condition that the total number of negative links is minimized.
In the continuum limit, negative links correspond, generally speaking
to singular fields. The data indicate that total volume of the defects
scales in physical units. We consider also correlator of negative
links. The correlator scales in physical units as well, within the
error bars. A new observation reported here is a strong anisotropy of
the correlator.}

\FullConference{XXIIIrd International Symposium on Lattice Field Theory\\
         25-30 July 2005\\
         Trinity College, Dublin, Ireland}

\begin{document}

\section{Introduction}

Understanding of non-perturbative fluctuations in the QCD vacuum has
been fast changing recently. The reason is the results of the lattice
simulations. In continuum theory one thinks mostly in terms of soft,
spherically symmetric excitations like  instantons. Lattice simulations, on the other hand,
indicate strongly that lower-dimensional defects are in fact crucial
for confinement, for a recent review see, e.g., \cite{vortices}.
Indeed, monopoles
are trajectories, or 1d defects, while the central vortices are
surfaces, or 2d defects.
Moreover, the 1d and 2d defects appear to be fine tuned, for review see
\cite{viz}. Namely, in case of the vortices the total area scales in
physical units \cite{vortices} while the  the total {\it
non-Abelian} action associated with the vortices is singular in the
limit of vanishing lattice spacing $a$ \cite{ft}:
\beq\label{area}
A_{vort}~\approx ~24(fm)^{-2}V_4~~,~~~ S_{vort}  ~\approx~0.54{A_{vort}\over a^2}~~.
\end{equation}
where $A_{vort}$ is the total area of the surfaces while $V_4$ is the
total volume of the lattice. Theoretically, the only way to explain the
observations (\ref{area}) is to assume that the surfaces possess also
ultraviolet divergent entropy which almost cancels the effect of
suppression due to the action (\ref{area}). Moreover, in case of
trajectories similar fine tuning is an indispensable part of the so
called polymer representation of field theory, see e.g. \cite{ambjorn}.

In case of surfaces, however, even imposing the fine tuning between
entropy and action `by hand' does not help much.
 In particular,
if one starts with the Goto-Nambu action and tunes it to the
entropy strings decay into what is called branched polymers
\cite{ambjorn}. The
branched polymers are effectively 1d structures.
The branched polymers are known to be relevant also to $Z(2)$ gauge theory
in 3d \cite{dotsenko} and in  4d \cite{kovalenko}.

The vortices observed in the vacuum state of gluodynamics are defined
and studied phenomenologically, through use of projected fields. How
then can one distinguish between `true' vortices (that is, Euclidean
strings) and branched polymers? The answer appears simple. Consider
minimal {\it three-dimensional} volume bound by the central vortices.
If the central vortices in the non-Abelian case are similar to the
central vortices of $Z(2)$ gauge theory then:
\beq\label{branched}(V_3)_{branched
~polymers}\sim\,a\cdot A_{vort}~~.\end{equation}
If, on the other hand, the central
vortices are true 2d dimensional defects then, generally speaking,
\beq\label{physical}
(V_3)_{strings}~\sim~ \Lambda^{-1}_{QCD}\cdot A_{vort}~~.
\end{equation}
In other words, the minimal 3d volume bound by the vortices is to scale
in the physical units. Results of measurements of the 3d volume were
reported first in \cite{volumes} and favor the possibility
(\ref{physical}).

 Further results on geometrical properties of the 3d
volume were obtained in \cite{novel}. In the subsequent sections we review
these results and their implications. Moreover, we add some further
preliminary results demonstrating anisotropy of the 3d volumes.

\section{Three-dimensional volumes}
\subsection{Scaling of 3d volumes}

Central vortices are defined as
unification of negative plaquettes in $Z(2)$ projection of the original
non-Abelian fields \cite{vortices}. Plaquettes evaluated in projected
fields are invariant under remaining (after the projection) $Z(2)$
gauge transformations. And in this sense the vortices themselves are
uniquely defined. The volume bound by the vortices is not uniquely
determined, on the other hand. The {\it minimal} volume bound by the
central vortices can be found by minimizing the number of negative
links. In other words, one fixes the
remaining $Z(2)$ gauge invariance by minimizing the total number of negative links.
In analogy with the $U(1)$ case,
the $Z(2)$ gauge considered  can be called $Z(2)$ Landau gauge fixing.  It turns out that the
volume occupied by the minimal number of negative links scales in the
physical units \cite{volumes}:
\beq\label{volume}
V_3~\approx~2(fm)^{-1}V_4~~,
\end{equation}
where $V_4$ is the total volume of the lattice.
As is explained in the Introduction,  result (\ref{volume}) implies that
the central vortices of gluodynamics are not branched
polymers but rather look as true  2d surfaces.

Let us also mention that there is no extra non-Abelian action
associated with the 3d volume under discussion \cite{volumes}. The
action is  the same as on average over the whole lattice. This is in
contrast with the case of monopoles and central vortices, which are
distinguished by an ultraviolet divergent non-Abelian action, see
\cite{anatomy} and \cite{ft}, respectively. However, this difference
can be readily understood theoretically along the lines of
argumentation presented in \cite{viz}.

\subsection{Removal of P-vortices}

Measurements of the 3d volumes are also relevant to appreciate the meaning of the
so called removal of P-vortices introduced in \cite{forcrand}. One determines first central projected values of the link
variables, $Z_{\mu}(x)$. And then modifies the original link matrices
$U_{\mu}(x)$ in the following way:
\beq\label{dd}
U_{\mu}(x)~\rightarrow~\tilde{U}_{\mu}(x),~~~~\tilde{U}_{\mu}(x)~\equiv~Z_{\mu}(x)U_{\mu}(x)~~.
\end{equation}
The effect of (\ref{dd}) is disappearance of confinement. Although this
is very impressive, there remains a question to be answered, how
serious is the damage to the original fields produced by an {\it ad
hoc} procedure (\ref{dd}). If one judges by the number of plaquettes
affected by (\ref{dd}) then the change affects a small fraction of the
whole lattice. Indeed, only plaquettes belonging to the P-vortices are
changing their sign and the probability of a given plaquette to belong
to P-vortices is small:
\beq\label{thetap}
\theta_{plaquette}~\sim~(a\cdot\Lambda_{QCD})^2~~,
\end{equation}
where $a$ is the lattice spacing and the probability (\ref{thetap})
tends to zero with $a\to 0$.

However, this cannot be a final answer to our question. Indeed, the
`vanishing of confinement' means that the value of the Wilson line for
a typical field configuration is changing its sign under (\ref{dd})
with a probability of order unit. Moreover, the Wilson line is 1d
object and P-vortices are 2d objects. Thus, generally speaking, they do
not intersect in d=4. Therefore, change of the sign of plaquettes
belonging to the P-vortices cannot be the reason for disappearance of
confinement under (\ref{dd}). The way out of the paradox is apparently
that we should follow change not only in plaquettes but in the
potentials (or links) as well and we come again to the 3d volumes considered above.

 Namely, the minimal number of links which are affected by (\ref{dd}) is
vanishing in the limit of $a\to 0$ as a 3d volume.
It vanishes, however, not so strongly as the number of plaquettes
belonging to P-vortices, see (\ref{thetap}). Note that the original version of
the removal of the P-vortices did not use the $Z(2)$ Landau gauge and
approximately half of the links were modified by (\ref{dd}).

\section{Correlator of negative links}
\subsection{Definitions}
As the next step, one can introduce correlator of negative links in the
$Z(2)$ Landau gauge. Since the total volume $V_3$ scales in the
physical units one may hope that the correlator scales as well
\footnote{Minimization of the number of negative links can be
considered as a discrete analog of minimization of $<(A_{\mu}^a)^2>$.
Although the latter vacuum expectation value is gauge dependent, its
minimal value may have a physical meaning \cite{a2}.}.
In the continuum
theory, if one imposes Landau gauge correlator of vector potentials is
described by a single form factor.
We are using $Z(2)$ Landau gauge and, at first sight, there is a single independent
form factor as well. However, the notion of a
negative link does not have meaning in the continuum. More precisely,
negative links correspond to singular fields, $A_{\mu}\sim 1/a$.
Therefore, we cannot rule out a priori more complicated dependencies on
the mutual orientation of the links and of the displacement $x$. We
will consider, therefore, the correlator of the negative links in its
generality and begin with corresponding definitions. Consider first
correlator of parallel links:
\beq \label{muravnonu}
G^{\parallel}_{\mu\nu}(x) = \langle Z_{0,\mu} Z_{x,\nu}\rangle,\, \mu =
\nu\end{equation}
Moreover, orientation of links and separation $x$ can
be either mutually transversal so that the scalar product of $x_{\rho}$
and of the unit vector in the $\mu$-direction vanishes, or
longitudinal so that the vector $x_{\rho}$ is directed along the $\mu$
direction as well.
One can also consider correlation of perpendicular links,
\begin{equation}
G^{\perp}_{\mu\nu}(x) = \langle Z_{0,\mu} Z_{x,\nu}\rangle,\,~~~~~ \mu \neq
\nu~~.
\end{equation}
Again, there are further sub-cases depending on the mutual
orientation of $x_{\rho}$ and links looking in the $\mu-,\nu-$
directions.

\subsection{Isotropic case}

Let us change variables,
$
\hat{Z}_{\mu}(x)~=~\{1,~if~Z_{\mu}(x)=-1;~~0~if~Z_{\mu}(x)=1\}~$.
Moreover, consider first the isotropic correlator:
\begin{equation}\label{isotropic}
G~\equiv~{1\over
N_r}\Sigma_{r<|x|<r+a/2}~\langle\hat{Z}_{\mu}(0),\hat{Z}_{\nu}(x)\rangle~~,
\end{equation}
where the summation runs over all links $Z_{\mu}(x)$ for $x$ lying in
the spherical layer $r<|x|<r+a/2$ and $N_r$ is the total number of
links in the layer.  At large distances the function
$G(r)$ can be fitted by a constant plus an exponent. The
corresponding mass turns to be close to the lowest glueball mass
\beq\label{glueball}
m~\approx~ (1.4-1.6)~GeV~~.
\end{equation}
Note also that measurements at finite temperature were performed
very recently as well \cite{reinhardt}.

\subsection{Anisotropy in the correlator of negative links}

A new point which we are reporting here is observation of a strong
anisotropy of the correlator of the negative links in the $Z(2)$ Landau
gauge. We will concentrate on the case $\mu=\nu$, see Eq.
(\ref{muravnonu}). 
Defining
\begin{equation}
F^{\parallel,\perp}~=~{G_{\mu\nu}^{\parallel,\perp}(x)\over
G_{\mu\nu}^{\parallel,\perp}(\infty)}~-~1,
\end{equation}
we observe that the correlators have {\it opposite signs}. Namely, the
correlation of the parallel negative links which we are considering is
positive if the displacement $x$ is perpendicular to the links and is
negative if the displacement vector is parallel to the links. Our
preliminary data are presented in Fig 1.
\begin{figure}[h]
\begin{minipage}[t]{0.5\textwidth}
\includegraphics[scale=0.4]{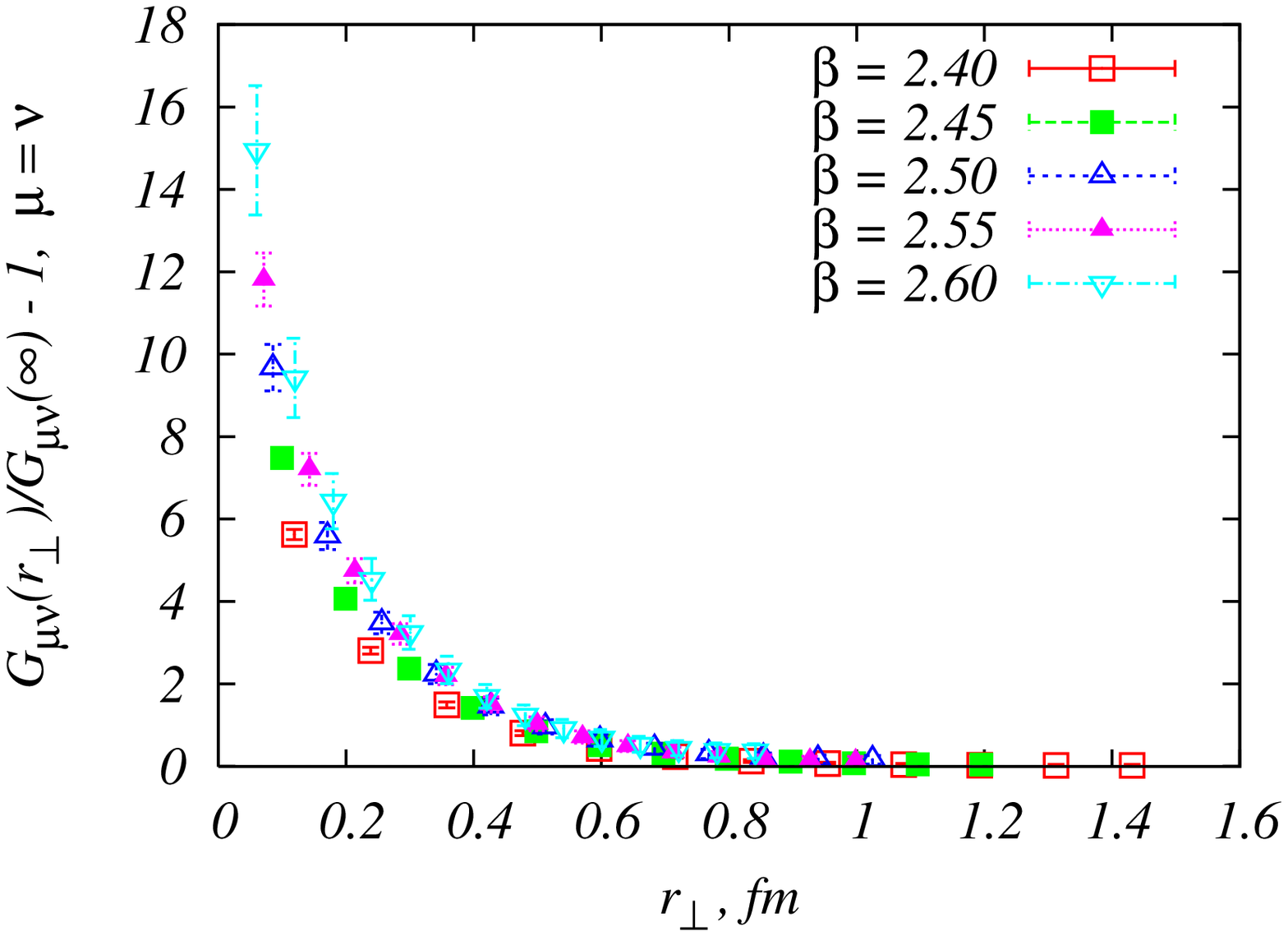} \\
(a)

\end{minipage}
\begin{minipage}[t]{0.5\textwidth}

\includegraphics[scale=0.4]{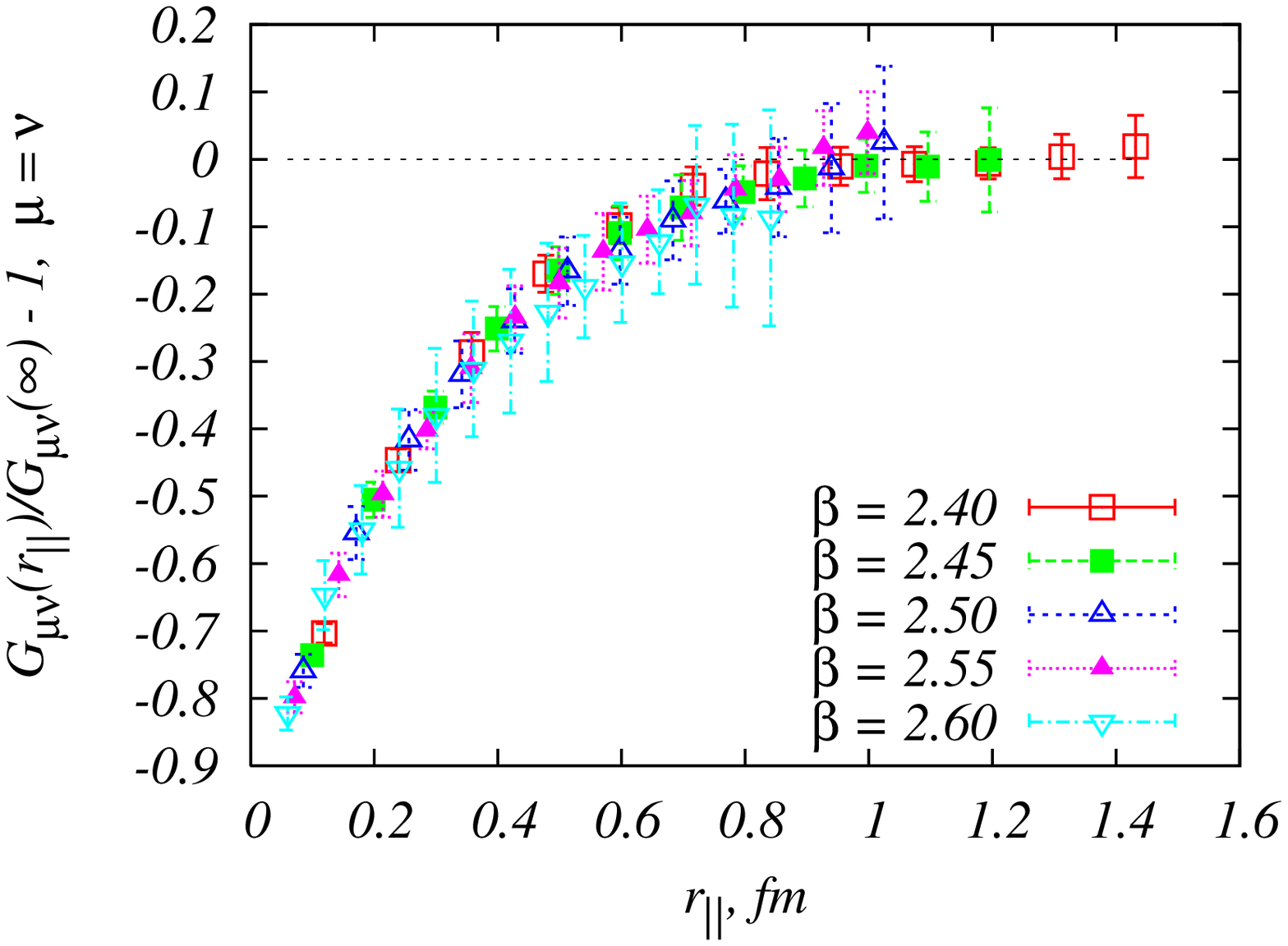}\\
(b)
\end{minipage}
\caption{Correlation of parallel links: (a) transversal correlation,
    $x=r_{\perp}$ is transversal to links, and (b) longitudinal,
    $x = r_{\parallel}$ is along the links. 
    \label{fig:imc_aniso}}
\end{figure}
The both correlators exhibit scaling. The mass values associated with the transversal correlator are
presented in Table \ref{tab:trans_mass}.
\begin{table}[h]
\begin{center}
\begin{tabular}{c|c|c|c|c}
\hline $\beta$ &$a,\,fm$&L &m, $fm^{-1}$ \\\hline  
$2.40$  &$0.1183$ &24   &$5.30\pm 0.10$\\
$2.45$  &$0.0996$ &24   &$5.35\pm 0.12$\\
$2.50$  &$0.0854$ &24   &$5.20\pm 0.15$\\
$2.55$  &$0.0713$ &28   &$5.40\pm 0.20$\\
$2.60$  &$0.0601$ &28   &$5.50\pm 0.10$ 
\end{tabular}
\caption{Mass parameter in the transversal correlator of parallel links.
    \label{tab:trans_mass}}
\end{center}
\end{table}

\subsection{Mass scales}

Since the  longitudinal correlator is negative,
the correlator of negative
links cannot be interpreted  as propagator of a physical degree of freedom. Rather, the
properties of the correlator reflect geometry of the
lower-dimensional defects. And at scale of $GeV^{-1}$ the
geometry is not isotropic (for a given field congfiguration).
The correlators scale in physical units and the
corresponding mass scales are having physical meaning. In particular,
the mass fitted above has the meaning of inverse typical size of the
3d volume in the transverse direction. In other words, existence of
lower-dimensional defects brings in mass scales which are not glueball
masses.

One can speculate, though, that at very large distances $x\gg
(GeV)^{-1}$ these masses are unobservable. Indeed, at such distances
the negative links will be separated by a few boundaries of the
percolating 3d volumes and the anisotropy should vanish. This guess is
partly confirmed by the observation that if, instead of going to limit
of very large $x$, one averages over the directions the correlation
length is indeed close to the inverse glueball mass, see Eq.
(\ref{glueball}).
\begin{acknowledgments}
The authors are grateful to J.~Greensite and to the members of ITEP
lattice group for stimulating discussions. This work was partially
supported by grants RFBR-04-02-16079, RFBR-05-02-16306a,
RFBR-05-02-17642, RFBR-DFG-436-RUS-113/739/0 and  EU Integrated
Infrastructure Initiative Hadron Physics (I3HP) under contract
RII3-CT-2004-506078.
\end{acknowledgments}

\end{document}